\documentclass[showpacs,aps]{revtex4}
\usepackage{graphicx}

\begin{document}

\title{Probing the center-vortex area law in d=3:  The role of inert
vortices}
\author{John M. Cornwall\footnote{Email:  Cornwall@physics.ucla.edu}}
\affiliation{Department of Physics and Astronomy, University of California, Los
Angeles CA 90095
\begin{center}
{\rm (Received January xx, 2006)}
\end{center}}

\begin{abstract}
\pacs{11.15.-q, 12.38.-t, 11.15.Tk   \hfill UCLA/06/TEP/01}

In center vortex theory, beyond the simplest picture of confinement several
conceptual problems arise that are the subject of this paper.  Recall that
confinement arises through configuration averaging of
phase factors associated with the gauge center group, raised to powers
depending on the total Gauss link number of a vortex ensemble with a given
Wilson loop. 
The simplest approach to confinement counts this  link number by counting the
number of vortices, considered in d=3 as infinitely-long closed self-avoiding
random walks
of fixed step length, piercing any surface spanning the Wilson loop.   
Problems
arise because a given vortex may pierce a given spanning surface several times
without  being linked or without contributing a non-trivial phase factor, or
it may contribute a non-trivial phase factor appropriate to a smaller number
of pierce points.      We
estimate the dilution factor $\alpha$, due to these inert or partially-inert
vortices, that reduces the
ratio of
fundamental string tension $K_F$ to the areal density $\rho$ of vortices  from
the ratio given by elementary approaches and find $\alpha =0.6\pm 0.1$.
Then we show how  inert vortices resolve the problem that the link number
of a given vortex-Wilson
loop configuration is the same for any spanning surface of whatever area, yet
a unique area (of a minimal surface) appears in the area law.    Third, we
discuss
semi-quantitatively a configuration of two distinct Wilson loops separated by
a variable distance, and show how inert vortices govern the transition
between two possible forms of the area law (one at small loop separation, the
other at large), and point out the different behaviors in $SU(2)$ and higher
groups, notably $SU(3)$.  The result is a finite-range Van der Waals force
between the two loops.  Finally, in a problem related to the double-loop
problem, we argue that the analogs of inert vortices do not affect
the fact that in the $SU(3)$ baryonic area law, the mesonic string tension
appears.

\end{abstract}

\maketitle

\section{Introduction}

To some extent, our understanding of area laws in confining gauge theories is
based on intuition and plausibility.  Certainly, there can be no doubt that,
in the fundamental representation of $SU(N)$,  the expectation value $\langle
W \rangle$ of the trace of a simple flat Wilson loop   $\Gamma$
involves the area $A$ of the flat surface spanning it and not the area of any
other spanning
surface:
\begin{equation}
\label{wloop}
W(\Gamma ) \equiv \frac{1}{N}Tr_F \exp [\oint_{\Gamma} dz_{\mu}A^{\mu}(z)];\;
\langle W \rangle =\exp [-K_FA].
\end{equation} 
[Here $K_F$ is the fundamental string tension, and we use imaginary
anti-Hermitean gauge potentials, incorporating the gauge coupling $g$ in them.]
But, at least in the center vortex picture of confinement, it is not always
easy to see how some of these plausible results follow from the basically
simple mechanism of confinement, based on linkages of vortices with Wilson
loops.  In this paper we discuss several confinement puzzles, all of them
connected by the theme of inert vortices.  By this we mean vortices that do
not link to Wilson loops in the usual way, but occupy space that could have
been occupied by truly-linked vortices.  For brevity we also use this term to
refer to partially-inert vortices that are linked, but with a smaller link
number than would naively be expected.  

We seek the effects of inert vortices only for large Wilson loops, those whose
length scales are all large compared to the fundamental gauge-theory length
$\lambda$.  As the loop scales approach $\lambda$ the effects of inert
vortices either disappear or are substantially modified.

  In the center vortex picture, the area law
arises through group-center phase factors raised to powers depending on the
Gauss link number of the vortex condensate with the Wilson loop.  This link
number can be calculated from the intersections of a vortex with any surface
spanning the Wilson loop.  To
characterize the condensate we will stick for simplicity to d=3, although
there is no real
qualitative difference between three and four dimensions.  In d=3, vortices
are closed stringlike tubes of chromomagnetic flux, a finite fraction of which
have infinite length,  and in d=4 they are closed 2-surfaces, whose description
raises complications.    In $d=3$ we model the vortices as closed
self-avoiding
infinite-length random walks on a cubical lattice of lattice length $\lambda$. 
This model is
similar in spirit, if not in implementation, to the $d=4$ models for $SU(2)$
and $SU(3)$ center vortices given by others \cite{engrein}, for use in lattice
computations.   It is also similar to the usual  identification of center
vortices as infinitesimally-thick objects, called P-vortices \cite{greensite},
that live on a lattice dual to the lattice where Wilson loops live.

Even for the archetypical example of a flat Wilson loop for $SU(2)$, inert
vortices are important.  In this simple case, there are many types of inert
vortices, including those that pierce the surface twice within a few
characteristic lengths $\lambda$, or that pierce it an odd number of times but
have link number less than the number of pierce points.     
The net effect of inert vortices for a simple flat Wilson loop is that the
density of linkage differs from the density of
piercing by a factor $\alpha$, which we call the dilution factor, lying between
zero and one.  This factor  is a rough but useful estimate of the various
ways in which vortices can be inert.  In Sec.~\ref{wloops} we estimate that
$\alpha$ lies in the
range 0.6$\pm$0.1. In the dilute gas approximation  (DGA), the usual
result for the $SU(2)$ fundamental string tension $K_F$ is $K_F=2\rho$;
dilution by inert vortices modifies this to  
\begin{equation}
\label{kvrho}
K_F=2\alpha \rho.
\end{equation}
[For $SU(3)$, the standard DGA gives $K_F=3\rho /2$; the diluted DGA is
$K_F=3\alpha_3 \rho /2$, as for $SU(2)$.  It may be that the $SU(3)$ dilution
factor $\alpha_3$ is not precisely the same numerically as it is for $SU(2)$,
but our estimates are not accurate enough to see much of a difference.  For
general $SU(N)$ there are in
principle as many dilution factors as vortex densities.]

It would, of course, be good to compare our estimates with lattice data. 
Unfortunately, it has turned out \cite{bfgo,forp,ltmr} to be rather difficult
to calculate the density $\rho$ on the lattice, for a number of reasons. Among
these are the dependence on gauge of the center-vortex location procedures;
effects of Gribov copies; and finite-size effects.  Ref. \cite{forp} states
that lattice artifacts are so important that these authors cannot really find
a reliable value for $\rho$.  However, \cite{ltmr} claims a value of
$K_F/\rho$ of 1.4, which taken literally might indicate a dilution factor
around 0.7, at least if the DGA is more or less correct. The best way to
attack the numerical estimation of dilution might be to simulate directly a
model of self-avoiding random walks, in the spirit of \cite{engrein}, rather
than to work with QCD itself.

A second major issue arises because link
numbers can be calculated (through Stokes' theorem) by counting intersections
of vortices with a surface spanning the Wilson loop, but any spanning surface
can be used for the link-number calculation, not just the surface that
ultimately appears in the area law.   So it is not clear what area is to be
used in the area law, nor even why there is a unique area.  We discuss these
issues in Sec.~\ref{unlink}, showing how unlinked vortices resolve the paradox
of two or more possible areas for a simple Wilson loop.

The third issue, elaborated in Sec.~\ref{2loops}, is an interesting variant on
the question of when vortices are linked or not.  This issue was raised before
at a qualitative level \cite{corn2004}.  We consider two identical
Wilson loops separated by a certain distance, and ask how the overall VEV of
these two loops depends on separation.  For $SU(2)$ this is a problem somewhat
like the corresponding soap-bubble problem, where there are (at least) two
minimal soap films that can appear for two wire frames close to one another. 
We can think of each loop as a $q\bar{q}$ meson, and our results indicate a
Van der Waals potential between the two loops, which breaks (as also happens
for soap bubbles) at a critical separation between the loops.  Similar but
more elaborate results hold for $SU(3)$.

Finally, in Sec.~\ref{baryon} we consider the baryonic area law for $SU(3)$.
This has been  explored in center vortex theory \cite{cornbar}, where it is
shown that the area law comes from three surfaces with quark world lines and a
central line as boundaries.  To some extent the baryonic area law seems to
have issues like those of the double Wilson loop in Sec.~\ref{2loops}; in
particular, there might appear to be correlations like the Van der Waals
potential of the two-loop problem that could modify the accepted baryonic area
law. 
However, we show that analogs of the inert vortices in the two-loop problem,
which are in fact not inert even though they are in the same geometry except
for orientation effects, do not
affect the fact that the linearly-rising potential for the three quarks has
precisely the mesonic string tension for each of its three sheets.   
Sec.~\ref{conc} contains a summary.

 Except for Secs.~\ref{principles}, \ref{wloops} we use the DGA approximation
to describe our results.  But it turns out to be convenient, for these two
other sections, to use a standard form \cite{greensite} of the area law that
contains the DGA as a limiting low-density case.

\section{\label{principles} Basic principles for the center-vortex area law}

The center-vortex picture for gauge group $SU(N)$  invokes  a vortex
condensate.  If $N>3$ there are $N-1$ types of vortices labeled by an integer
$k,\;1\leq k\leq N-1$ that gives the vortex magnetic flux in units of $2\pi
/N$.  A vortex labeled $k$ is the antivortex of the vortex labeled $N-k$. 
These vortices are characterized by an areal density
 $\rho_k=\rho_{N-k}$ for each vortex type.  By this areal density we mean (in
all dimensions, not just $d=3$) that the average number of $k$-vortices that
pierce any flat surface of area $A$ is $\rho_kA$.  There is not much
theoretical insight into the
values of these different densities for $SU(N)$ with $N\geq 3$.  However, for
$SU(2)$ and $SU(3)$ there is only one density, which we term $\rho$, that sets
the scale for the string tension (in $SU(3)$ there are two types of vortices,
but one is the antivortex of the other and they have the same densities).   In
this paper we will only consider gauge groups $SU(2)$ and $SU(3)$. 

 A condensate of vortices can form only if a finite fraction of them has
essentially unbounded length (or an unbounded number of steps in the random
walks describing the vortices).  Only such
vortices, long compared to any Wilson loop scale, can contribute to the area
law.  These can be linked or not,
depending on the circumstances we encounter.  Finite-length vortices are
therefore inert, in our terminology, and will be accounted for by a
renormalization factor that we will not attempt to calculate here.

Consider now a simple flat Wilson loop and the flat surface spanning it.  This
surface is taken to lie in a plan of the lattice dual to the vortex lattice
and is divided into squares of this dual lattice; we  call these
 $\lambda$-squares.   Any such square is pierced by a single vortex with
probability $p$.  This probability is related to the areal density of the
vortex condensate by
\begin{equation}
\label{prob}
p=\rho \lambda^2;
\end{equation}
 the probability that a square is unoccupied is $\bar{p}=1-p$.    As on the
lattice, $p$ can be extracted 
\cite{greensite,bfgo,forp,ltmr} from the VEV of a square Wilson loop one
lattice unit on a side.  Denote this VEV as $\langle W(1\times 1)\rangle$;
then for $SU(2)$
\begin{equation}
\label{pdef}
\langle W(1\times 1)\rangle =\bar{p}-p;\;\;p=\frac{1}{2}[1-\langle W(1\times
1)\rangle].
\end{equation} 
The only difference from the lattice definition is that in lattice computations
the length scale for the Wilson loop is not a physical quantity $\lambda$ but
a lattice spacing, and the lattice version of $p$ must be scaled via the
renormalization group to find a physical probability, such as used in the
present model [Eq.~(\ref{prob})].   Note that $p$ is bounded by 1/2, since  
$\langle W(1\times 1)\rangle$ lies between 0 and 1.  Moreover, note that the
probability $p$ as derived from Eq.~(\ref{pdef}) is not subject to dilution,
which applies only to large Wilson loops.

The assumption that vortices in different $\lambda$-squares are statistically
independent leads to the standard argument for center-vortex confinement,
which ignores inert
vortices.  The confining area law (discarding perimeter effects) for a
Wilson loop follows from an ensemble average of center-group elements.   Each
of these is of the form $\prod_i Z_i^{Lk_i}$, where $Z_i$ is an element of the
center of the gauge group as specified by the properties of the $i^{th}$
vortex (and the group representation of the loop itself).  For example, for the
fundamental Wilson loop in $SU(2)$ the only non-trivial element of the center
has $Z_i=-1$.  In the above product, $Lk_i$
is the Gauss linking number of this vortex with the Wilson loop.  The Gauss
linking number, a topological quantity, can be written as an intersection
number of the vortex and any surface spanning the Wilson loop; its (integral)
value is independent of the choice of surface.   In the $SU(2)$ case the
necessary average is
\begin{equation}
\label{center}
\langle \exp [i\pi \sum_i Lk_i]\rangle.
\end{equation}

Aside from the assumption of independent $\lambda$-squares, the critical
assumption  for expressing confinement in the
center-vortex picture is that $p$ is the probability that a vortex is
actually linked once to a flat Wilson loop.  When an odd number of vortices is
linked once, the Wilson loop has value -1 and when an even number is linked,
the value is +1.   If the assumption is true, the area law follows from
multiplying the probabilities $\bar{p}-p$ of Eq.~(\ref{pdef}) for all the
$\lambda$-squares of the spanning surface.

Another useful way of expressing this area law is to write out the
combinatorics for vortex occupancy of $N_S$ sites of a surface spanning a
Wilson loop $\Gamma$:  
\begin{equation}
\label{vev}
\langle W_{\Gamma}\rangle =
\bar{p}^{N_S}-N_S\bar{p}^{N_S-1}p+\frac{N_S(N_S-1)}{2}\bar{p}^{N_S-2}p^2+\dots
=[\bar{p}-p]^{N_S}=[1-2p]^{A_S/\lambda^2}.
\end{equation}
Here the number $N_S\gg 1$ of sites on a given spanning surface $S$ is
$N_S=A_S/\lambda^2$, where the surface has area $A_S$, and each term represents
the number of ways of arranging empty and once-filled $\lambda$-squares.

Next we need to modify the area law for dilution, which arises from several
factors:  If a vortex penetrates an even number
$2N_p$ of
times, there are $N_p$ sites that lead to unit phase factor in the Wilson-loop
VEV, although these sites are occupied.  In effect,
the vortices filling such sites are inert (although they may, strictly
speaking, be linked to the Wilson loop topologically).  Similarly, if a vortex
penetrates an odd number (greater than 1) of times  it is linked and gives a
non-trivial phase factor, but three or more sites are occupied, rather than
the single site assumed when we related the string tension and the piercing
probability as in Eq.~(\ref{vev}). We will argue in Sec.~\ref{alpha} that the
diluted form of the standard equation (\ref{vev}) is
\begin{equation}
\label{dilform}
\langle W_{\Gamma}\rangle = [1-2\alpha p]^{A_S/\lambda^2},
\end{equation}
yielding a string tension
\begin{equation}
\label{dilstring}
K_F=-\frac{1}{\lambda^2}\ln (1-2\alpha p).
\end{equation}
The DGA approximation  is the small-$p$ limit of either Eq.~(\ref{vev})
(undiluted) or of Eq.~(\ref{dilstring}) [diluted; see 
Eq.~(\ref{kvrho})].  Note that while the effect of dilution on the probability
$p$ is simply to renormalize it, the effect of dilution on a dimensionless
quantity such as $K_F/\rho$ cannot be characterized as a renormalization of a
dimensionful quantity such as the density $\rho$ or the $\lambda$-square area
$\lambda^2$:
\begin{equation}
\label{nonrenorm}
\frac{K_F}{\rho}= \frac{-\alpha}{x}\ln (1-2x)|_{x=\alpha \rho \lambda^2}.
\end{equation}

It seems very difficult to resolve  dilution problems completely by analytic
methods; the best we can do is to give a semi-quantitative discussion of how 
these factors renormalize downward the linkage probability from $p$.

\section{\label{wloops} The dilution factor for a flat surface}

In this section, dealing with the gauge group $SU(2)$, we distinguish between
the previously-introduced probability $p$ that a vortex pierces a
$\lambda$-square, thus contributing unit link number,  and the  density of link
number per $\lambda$-square, which is what we really need.  This density is
reduced by inert-vortex effects.  We attempt to capture, in some mean-field
sense, these effects approximately by introducing a dilution factor $\alpha$
that effectively reduces the pierce probability $p$ to $\alpha p$.

 We begin by classifying various ways in which vortices can pierce a spanning
surface yet not be linked (in the sense that they are associated with a
trivial phase factor), or are linked but are to be associated with a reduced
density of linkage numbers.  For brevity we refer to all these as inert
vortices.

\subsection{\label{types} Types of inert vortices}

 It is useful to distinguish three types of inert vortices; only Types II and
III need detailed discussion.
Type I vortices have finite length, and correspond in some sense to localized
particles.  The vortex condensate may have some of these, but they cannot
explain confinement, since for large Wilson loops those that are linked
contribute only to sub-area effects such as perimeter terms in the VEV.  
Presumably the effect of
such vortices is essentially to renormalize the areal density of vortices of
unbounded length.  We will not discuss such finite-length vortices any
further, so from now on we are only concerned with the issue of vortices much
longer than any Wilson loop scale, and the extent to which these are or are
not inert.

 The distinction between the remaining two types of long vortices is this. 
Type II vortices exhibit what we will call local return, by which we mean that
a vortex, however long, penetrating a localized flat surface has its highest
probability of returning to that surface after only a few more steps of the
random
walk.  This probability is not to be confused with the probability that a
random walk will be near to where it has been  after a large number $N_s$ of
steps; this probability decreases like $N_s^{-d/2}$, where $d$ is the
dimension of space-time (see the Appendix).    For the present section, dealing
with a flat spanning surface, this type is important, but once the
renormalization of $p$ has been made, they are not very important in
the further applications of Secs. \ref{unlink}, \ref{2loops}.

Finally, a Type III vortex is one which, having penetrated a surface,
penetrates it a second time with high probability after a large number of
steps.  This can only be true, in view of the above remarks, if the surface is
of a special type, including those used in Secs. \ref{unlink}, \ref{2loops},
and \ref{baryon} such as a
closed surface or one that has curvature radii comparable to its length
scales.  An infinite-length vortex must penetrate a  closed
surface at least twice.    

A qualitative model of Type III effects might be to assume that after
renormalization
for the effects of Types I and II vortices, the remaining  effects are seen
for vortices composed of infinitely-long straight lines intersecting surfaces
compounded of flat segments. The rationale for the straight-line vortices
is that the local returns have been accounted for by renormalization for Type
II vortices.  In this model no vortex can return locally to
a single surface element and so explicit Type II vortices are missing; more
than one flat surface segment must be involved, and these
must form a non-planar surface. We will not use this model in the present
paper.

\begin{figure}
\includegraphics[height=4in]{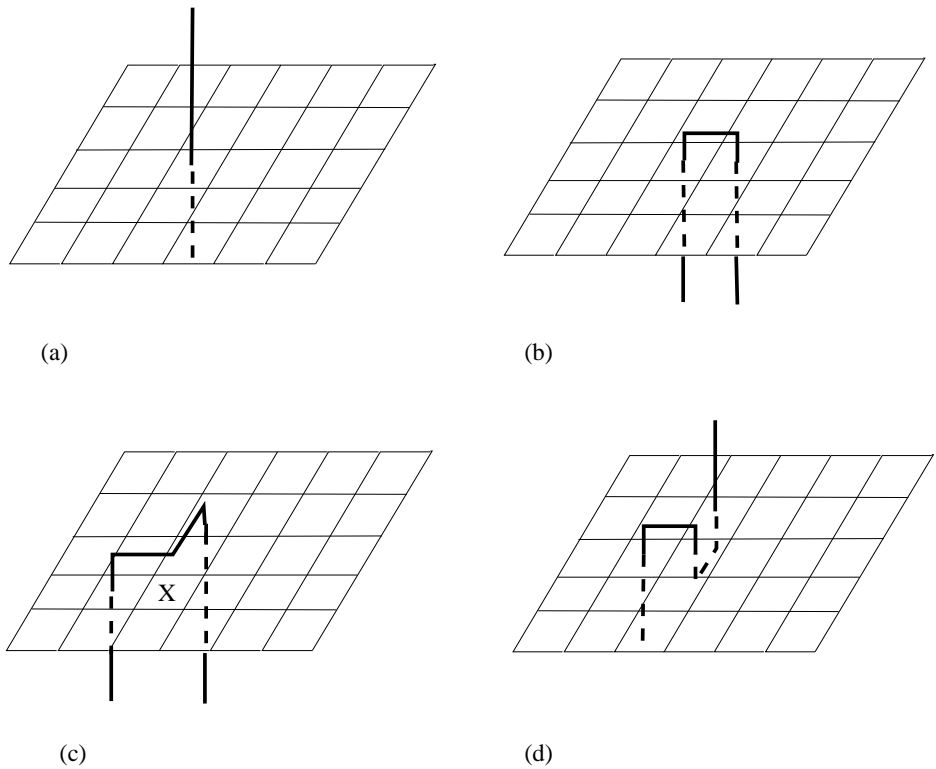}
\caption{\label{fig1}  In these figures, the surface being pierced lies in a 
plane on a lattice (shown) dual to the vortex lattice (not shown).  (a) 
Simple piercing by a vortex (thick line).  (b)  An inert vortex, penetrating
the surface with two more steps after the first piercing.  (c)  An inert
vortex that has rendered the square labeled X inaccessible to another vortex. 
(d)  A partially-inert vortex, using up three $\lambda$-squares for a single
linking.}
\end{figure}

Among the Type II vortices, the only ones to be considered in this section,
there is
essentially an infinite number of subtypes.  Several of them are shown in
Fig.~\ref{fig1}.

The first, Fig.~\ref{fig1}(a),  shows the standard penetration assumed in the
usual area-law formula of Eq.~(\ref{vev}):  Unit link number associated with a
single piercing.  Fig.~\ref{fig1}(b) shows an inert vortex (zero link number)
produced by two additional steps from the first piercing.  Fig.~\ref{fig1}(c)
shows another inert vortex, using three additional steps, that makes the
square labeled X inaccessible for piercing by another vortex, because of the
mutual-avoidance requirement. Fig.~\ref{fig1}(d) shows a vortex with four
extra steps that is linked, but has pierced three $\lambda$-squares, thereby
again reducing the $\lambda$-squares available for more vortices.  Note that
in every case shown in the figure except for Fig.~\ref{fig1}(a)  the link
number is reduced relative to the piercing number. 

There are other ways for vortices to be inert, for example, a vortex may have a
link number that is a multiple of $N$ for $SU(N)$ with $N>2$.   Related effects
take place in the baryonic area law for $SU(3)$ (Sec.~\ref{baryon}).

\subsection{ \label{alpha} Definition of $ \alpha$}

The flat surface is divided, as before, into $N_S$ $\lambda$-squares.  Of
these, on the average $pN_S$ are pierced once by a vortex, and of the pierced
squares, a
fraction $\alpha$ on the average contribute to the area-law formula with a
minus sign (in
$SU(2)$). 

We use a simple statistical model, ignoring shape effects and certain
correlations and assuming that dilution is statistically independent of
piercing.  It is easy to see that the coefficient of 
\begin{equation}
\label{alphaexp}
\bar{p}^{K_1}(\alpha p)^{K_2}[(1-\alpha )p]^{K_3}\;\;[\sum K_i=N_S]
\end{equation} 
in the formal expansion of
\begin{equation}
\label{formexp}
1=[\bar{p}+\alpha p+ (1-\alpha )p]^{N_S}
\end{equation}
is the statistical weight for a configuration with $K_2+K_3$ pierced
$\lambda$-squares, of which $K_2$ are going to give a minus sign in the Wilson
loop VEV.  We then have
\begin{equation}
\label{alphaloop}
\langle W \rangle =[\bar{p}-\alpha p+ (1-\alpha )p]^{N_S}=[1-2\alpha p]^{N_S},
\end{equation}
showing that, as previously specified, $\alpha$ simply renormalizes $p$.
For the DGA formula for the Wilson-loop VEV we find 
\begin{equation}
\label{renorm}
\langle W_{\Gamma}\rangle = 
\exp[-2\alpha \rho A_S],
\end{equation}
in which the string tension is renormalized by the factor $\alpha$ from its
previous value.  The DGA is perhaps made more plausible by dilution.     

Not all the effects of inert vortices can be captured by the simple dilution
factor defined above.  Vortices are correlated with each other through
self-avoidance, as in Fig.~\ref{fig1}(c), and the specific geometry of the
portion of a random vortex walk that penetrates the spanning surface more than
once can matter.  Unfortunately, even simpler problems cannot be solved
analytically.  For example, the statistics of co-existing but mutually- and
self-avoiding monomers and dimers (see \cite{nemir} for a mean-field approach
and earlier references), has no exact solution (except in the limit of
close-packed dimers \cite{tempfish}). This is because dimers and higher
multimers do not obey simple ({\em e. g.},
multinomial) statistics.  One can appreciate this from the observation that on
an empty lattice of $N_S$ sites a dimer can be laid down in $2N_S$ ways, but
only $N_S/2$ self-avoiding dimers can be put down in total.  (For monomers, of
course, these two numbers are the same, namely $N_S$.)

In the inert-vortex problem there are, in principle, a huge number of
multimers, of various shapes and sizes, that should be accounted for.  Rather
than attempting some elaborate generalization of the monomer-dimer problem, we
proceed as follows.   

\subsection{\label{findalpha}  Estimating the dilution factor}

The problem is to estimate the probability that a self-avoiding random walk of
infinite length, having pierced the Wilson surface (which we call $W$) once,
pierces it again one or more times.
 We need two different types of probabilities for this problem.  The first is
the standard probability density $p_3(N;\vec{m})$ that a random walk on the
$d=3$ lattice is at the lattice point $\vec{m}$ after exactly $N$ steps.  If
the difficult restriction of self-avoidance is dropped, this is given by (see,
{\em e. g.}, \cite{montroll})
\begin{equation}
\label{3dprob}
p_3(N;\vec{m})=\frac{1}{(2\pi )^3 3^N}    [\prod_{j=1}^3\int_0^{2\pi}d\theta_j]
[\cos \theta_1+\cos \theta_2+\cos \theta_3]^N\exp [i\vec{\theta}\cdot\vec{m}].
\end{equation} 
This is normalized so that 
\begin{equation}
\label{probnorm}
p_3(N=0;\vec{m})=\delta_{\vec{m},\vec{0}}
\end{equation}
 and the sum over all $\vec{m}$ of this density yields unity. In the limit of
large $N$ and components of $\vec{m}$ (which is an integer-valued vector)
$p_3(N;\vec{m})$ has the usual Gaussian form.  This is not the probability of
real interest, although it can be used in certain circumstances to find the
probability that we need.  We will call the one that we do need
$q_3(N;W;\vec{m})$, the probability density that a
random walk piercing at the origin re-pierces for the first time at $\vec{m}$
using exactly
$N$ steps, with the vector $\vec{m}$ restricted to lie in the lattice plane
that is nearest neighbor to the Wilson surface.   
The probability density $p_3(N;W;\vec{m})$ is simply $p_3(N;\vec{m})$ with
$\vec{m}$ restricted to the Wilson surface.  Our
interest is in probabilities summed over the Wilson surface, so we define
\begin{equation}
\label{sumprob}
p_3(N;W)=\sum_{\vec{m}\in W}p_3(N;W;\vec{m});\;\;q_3(N;W)=\sum_{\vec{m}\in
W}q_3(N;W;\vec{m}).
\end{equation} 

By the standard rules of probability \cite{feller}
\begin{equation} 
\label{compprob}
p_3(N;W)=\delta_{N,0}+\sum_2^{N}q_3(J;W)p_3(N-J;W)
\end{equation} 
which says that the random walk, having pierced the surface for the first time
after $J$ steps, may penetrate it many times again before ending on the
surface after $N$ steps.  [Note that it takes $N=2$ additional steps of the
random walk, at minimum, to re-pierce the surface, given that these steps are
counted as starting with the first step after the first piercing.]
  If the total number of piercings is odd (even) the vortex is linked
(unlinked). The case of no further intersections after the first piercing is
to be included; the standard area-law model of Sec.~\ref{principles} is
equivalent to assuming that this probability is unity, and all other
probabilities are zero.  This is far from the case, as we will see.  

Eq.~(\ref{compprob}) is easily solved in terms of generating functions
\begin{equation}
\label{genfuncdef}
P_3(s;W)=\sum_0 p_3(N;W)s^N;\;\;Q_3(s;W)=\sum_2q_3(N;W)s^N.
\end{equation} 
We have
\begin{equation}
\label{pvsq}
P_3(s;W)=1+P_3(s;W)Q_3(s;W);\;\;Q_3(s;W)=1-\frac{1}{P_3(s;W)}.
\end{equation}
Note that we may express the elementary solution for $P_3(s;W)$, which is 
\begin{equation}
\label{peqn}
P_3(s;W)=\frac{1}{1-Q_3(s;W)},
\end{equation}
in a suggestive way for the original probabilities:
\begin{equation}
\label{qeqn}
p_3(N;W)=1+\sum q_3(K;W) + [\sum q_3(K;W)]^2+ \dots 
\end{equation}
showing how the probability $p_3$ is compounded from probabilities of first
return [see, {\em e. g.}, Fig.~\ref{fig1}(d)].

The final probability of interest is the probability that the random walk ever
re-pierces the surface; this is clearly given by the sum of all the $q_3$, or
by $Q_3(s=1;W)$. 

Equation (\ref{pvsq}) also holds in certain other problems, notably the
gambler's ruin problem (described in the Appendix), which asks for the
probability that a $d=1$ random walk starting at the origin ever returns to it. 
In
this problem backtracking (re-tracing the last step) is allowed, and (as we
review in the Appendix) it is
straightforward to show that the corresponding probability of ever returning
is exactly unity.  It is also unity in two dimensions.   However, our problem
differs from the gambler's ruin problem in two essential ways:  No
backtracking is allowed, and ultimately the problem becomes three-dimensional,
when the radius of gyration $N^{1/2}$ of the random walk becomes large
compared to the size scale $L$ of the Wilson loop.  [Actually, the radius of
gyration for self-avoiding walks has an exponent somewhat different from 1/2,
but we ignore that complication here.]  It has long been known (as reviewed in
\cite{montroll}) that the probability of return to the origin in $d=3$ is
finite, with a value of approximately 0.34.  Furthermore, self-avoidance
completely changes the problem; for example, a self-avoiding ``random" walk in
$d=1$ has probability zero of ever returning to any site it has reached.

Let $\mathcal{P} $ be the probability that a vortex, having penetrated the
Wilson
surface once, never penetrates it again.  Then, with $Q_3(1;W)$   as the
probability that it ever penetrates the surface again, 
\begin{equation}
\label{wpsoln}
\mathcal{P} = 1 - Q_3(1;W).
\end{equation}
 Or one may write, as in Eq.~(\ref{qeqn}), a probability sum rule saying that
the sum of probabilities of piercing exactly once, exactly twice, etc., must
be unity:
\begin{equation}
\label{feller2}
 1=\mathcal{P} \{1+\sum q_3(K;W) + [\sum q_3(K;W)]^2+ \dots \}.
\end{equation}
As before, the sum over $K$ begins with 2 for self-avoiding random walks.

It is now necessary to estimate $\mathcal{P}$, or equivalently $Q_3(1;W)$. 
Unfortunately, this is not a straightforward matter when there is
self-avoidance (see the Appendix for a brief review of some of the well-known
analysis when this condition is not imposed).  One way to proceed is simply to
count
the number of self-avoiding paths going from an original piercing to another
piercing as a function of their step length, and look for ways of partially
re-summing the results.  Our proposal, given below, is more accurate for random
walks that do not backtrack than for true self-avoiding walks; the difference
is that a non-backtracking walk may violate the condition of self-avoidance by
looping back on itself.  We  propose the following  expression for
non-backtracking walks, valid for $J\geq 2$:
\begin{equation}
\label{smallqeqn}
q_3(J;W)=(\frac{4}{25})\frac{1}{2}[(\frac{3}{5}+\frac{1}{5})^{J-2}+
(\frac{3}{5}-\frac{1}{5})^{J-2}].
\end{equation}
The corresponding generating function, approximately valid for self-avoiding
walks, is
\begin{equation}
\label{qapproxgen}
Q_3(s;W)=(\frac{2}{25})[\frac{1}{1-(4s/5)}+\frac{1}{1-(2s/5)}].
\end{equation}
The explanation of the terms is as follows.  The factor 4/25 is the probability
for Fig.~\ref{fig1}(b), for $J=2$.  For a non-backtracking walk in $d=3$ there
are 5 possible choices to add a new step to the random walk.  The horizontal
step in this figure has probability 4/5 (note that we are summing over all
possible sites), and the next, vertical, step has probability 1/5.  The
appearance at larger $J$ of 3/5 in the formula expresses the probability that a
non-backtracking  random walk will take its next step horizontally
(with respect to the Wilson surface), and 1/5 is the probability of a vertical
step in one particular direction (up or down).  [The restriction to even
powers of 1/5 is easily understood by drawing a few figures, in the style of
Fig.~\ref{fig1}.]  So a random walk once started in a horizontal plane has a
three times larger probability of staying in that plane than of  moving to a
plane higher above the Wilson surface.   It is easy to check the combinatorics
of the low-order powers of 1/5 in Eq.~(\ref{smallqeqn}), and also the highest
powers, for example, the last term (4/25)$(1/5)^{J-2}$.  This is the
probability that the random walk goes as high as
possible, takes one horizontal step, and then returns straight down to the
Wilson surface.  This can only happen for even $J$, and one easily sees that
the maximum attainable height is $J/2$.    We have compared the approximation
of Eq.~(\ref{smallqeqn}) with explicit counting of self-avoiding walks through
$J=8$ and find that the difference between self-avoiding and non-backtracking
is acceptably small.   
 
The difference between non-backtracking walks and self-avoiding walks first
appears at $N=5$, where there are 144 non-backtracking walks but only 128
self-avoiding walks.  This is about an 11\% error, but because the erroneous
contribution to the $N=5$ walks is 16/3125 $\simeq$ 0.0052, out of a total of
about 0.533, the error in the final result from $N=5$ terms is less than 1\%. 
In any case, to deal with true self-avoidance rather than just
non-backtracking raises the same issues as having a condensate of
mutually-avoiding vortices, and we do not discuss that issue in any detail in
this paper.

We find an approximation to $Q_3(1;W)$ by summing over all $J$ in
Eq.~(\ref{smallqeqn}), with the result
\begin{equation}
\label{largeqapprox}
Q_3(1;W)\simeq (\frac{4}{25})\times \frac{1}{2}[\frac{1}{1-(4/5)}+
\frac{1}{1-(2/5)}]=\frac{8}{15}\simeq 0.533.
\end{equation}

At this point one might guess that the dilution factor $\alpha$ should be equal
to the probability $\mathcal{P}$ of never returning .  But this is not
quite right; we really need the probability of being linked or unlinked. To do
this, one should separately find the probabilities that there is an even
or odd number of piercings, as expressed by separating the even and odd powers
of $Q_3(1)$ in Eq.~(\ref{feller2}).  We use $\mathcal{P}_L$ for the probability
that a
vortex is linked, and $\mathcal{P}_U$ for the probability that it is unlinked;
these
are defined by
\begin{equation}
\label{linkunlink}
\mathcal{P}_L=\mathcal{P}\{1+
\sum_{even}Q_3(1)^N\}=\frac{1}{1+Q_3(1)};\;\;\mathcal{P}_U=\mathcal{P}
\sum_{odd}Q_3(1)^N
=\frac{Q_3(1)}{1+Q_3(1)}.
\end{equation}
Now it seems that $\alpha =\mathcal{P}_L$.
If  we to use our approximation $Q_3(1)=0.53$ we would find the link
probability (dilution factor) to be
about 0.65, somewhat larger because we are counting three, five, $\dots$
piercings as well as a single piercing.

Eq.~(\ref{linkunlink}) is not quite right either,
because while $\mathcal{P}_L$ expresses the probability of linkage, when (say)
three
sites are used for the link
instead of one, there is dilution that must be accounted for.  We account for
this by calculating a sum of weighted probabilities, where the weight for the
appearance of $K$ powers of $Q_3(1)$ in Eq.~(\ref{linkunlink}) is the inverse
of the number of sites occupied by the linked vortex, or $(K+1)^{-1}$. For
example, when $K=2$ three $\lambda$-squares are pierced, as shown in
Fig.~\ref{fig1}(d).   This result for
$\alpha$, when modified by the original piercing probability $p$, should give
something like the link number per unit area, in effect increasing the size of
$\lambda$-squares to account dilution.     The weighted  sum for
$\mathcal{P}_L$ gives
another estimate for $\alpha$:
\begin{equation}
\label{newalpha}
\alpha \simeq
\mathcal{P}\{\sum_{even}\frac{Q_3(1)^N}{N+1}\}=\frac{1-Q_3(1)}{2Q_3(1)}
\ln [\frac{1+Q_3(1)}{1-Q_3(1)}].
\end{equation} 
For $Q_3(1)=0.53$ this gives $\alpha = 0.52$.

It is plausible that $Q_3(1;W)$ lies between the no-self-avoidance $d=3$ value
of 0.34 and the $d=1$ value of 1.  In that case, the estimate of $\alpha$ with
no inverse-site weighting [Eq.~(\ref{linkunlink}] gives a finite value even in
the $d=1$ limit, where we find
$\mathcal{P}_L=\mathcal{P}_U=1/2$, so the dilution factor is 1/2.  This is a
singular limit,
because the original probability $\mathcal{P}$ used to construct
$\mathcal{P}_L,\mathcal{P}_U$
vanishes, but this is cancelled by a singularity in the sum over $Q$s.  It is
not surprising that the limiting probabilities are each 1/2, since there is no
way of distinguishing an even number of piercings from an odd number.  The
value of $\alpha$ from this equation for the $d=3$ value $Q_3(1;W)=0.34$ is
about 0.75.  Eq.~(\ref{newalpha}), with inverse-site weighting, gives 0.75 for 
 $Q_3(1;W)=0.34$ and 0 for $Q_3(1;W)=1$.  Presumably this latter case is
unrealistic, because at some point the radius of gyration of the random walk
is large compared to the Wilson loop scale $L$ and the problem really is
three-dimensional; in any case, the $d=1$ self-avoiding case is completely
opposite to the no-self-avoidance case.

To be more accurate in estimating $\alpha$, one would need to account for
mutual avoidance effects such as shown in Fig.~\ref{fig1}(c), and be more
precise about weighting various random-walk configurations.  We will not
attempt that here, and close by saying that our estimates for the dilution
factor are consistent with $\alpha=0.6\pm .1$. 

 In the following sections we  look for further manifestations of inert
vortices of Type III, going beyond the local effects that led to the dilution
factor.  But these local effects still occur, and so everywhere in our
arguments the original probability $p$ should be replaced by the diluted
probabilities
\begin{equation}
\label{diluteprob}
\hat{p}\equiv \alpha p;\;\;\bar{p}\equiv 1-\hat{p}.
\end{equation}

\section{\label{unlink} Inert vortices and the spanning surface}

Now we return to the problem that the spanning surface used for counting link
numbers as piercings is arbitrary, yet there surely is a unique area in the
area law.
The simplest possible case, where there is no doubt as to the answer, is that
of a flat Wilson loop, and one can get the right area law by using the flat
spanning surface in Stokes' theorem.  Yet one should also get the right answer
by using any other surface.  How does this come about?

\begin{figure}
\includegraphics[height=3in]{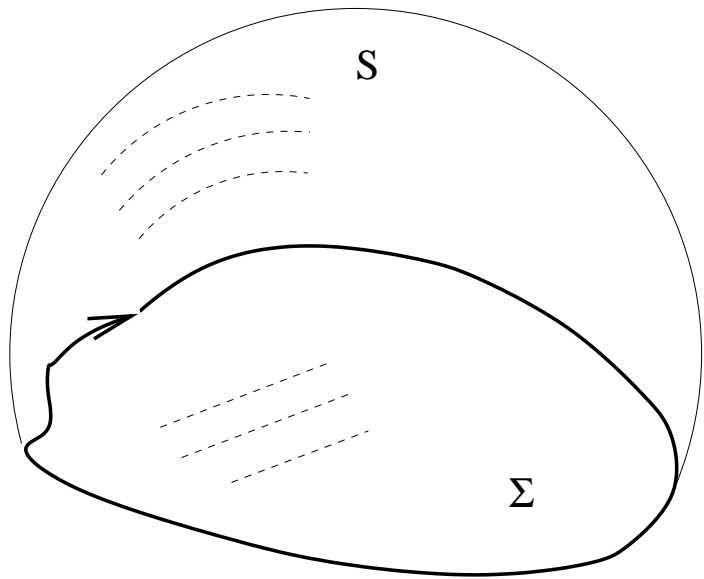}
\caption{\label{fig2} A flat Wilson loop, and two spanning surfaces, labeled
$\Sigma$ and $S$.} 
\end{figure}

Fig.~(\ref{fig2}) shows such a flat Wilson loop, with two spanning surfaces. 
The first, labeled $\Sigma$, is flat and is correct.  The second is labeled
$S$.  We
choose orientations so that the combined surface $\Sigma + S$ is oriented.  Of
course,
this surface is closed.

We must now improve upon the techniques outlined in Sec.~\ref{unlink}, to
account for inert vortices of type III.   The calculation of the area law
based on the flat surface $\Sigma$ needs no change.  But what if we instead
wished to calculate the area law based on surface $S$?   Here there is extra
dilution.   The vortices linked to surface $S$ are still those linked to
surface $\Sigma$, which must pass
through
surface $\Sigma$ and surface $S$ once each, but there are also inert vortices,
which pass
through surface $S$ twice and surface $\Sigma$ not at all.  [We will not
account for vortices linking three, five, $\dots$ times, most of which are
included in the dilution factor.]  We denote the number of linked vortices by
$N_L$ and the number
of inert vortices by $N_I/2$.  The factor of 1/2 in the latter definition
simply reflects the fact that every inert vortex pierces surface $S$ twice,
so that $N_I$ is the total number of pierce points of inert vortices.  It
is not convenient to introduce this factor of 1/2 for linked vortices, because  
the diluted pierce probability $\hat{p}$, introduced above, is related
to  the number $N_L$ of vortices linked to surface $\Sigma$ by
\begin{equation}
\label{nl1}
N_L=\frac{\hat{p}A_{\Sigma}}{\lambda^2}
\end{equation}
where $A_{\Sigma}$ is the area of surface $\Sigma$.  That is, $N_L$ is the
number of pierce points of vortices on surface $\Sigma$.  

The total number of vortex piercings of  the combined surface $\Sigma + S$,
denoted $N_{tot}$, is
\begin{equation}
\label{1plus2}
N_{tot} = 2N_L+N_I, 
\end{equation}
with the two arising because linked vortices penetrate both surface $S$ and
surface $\Sigma$.  
 By hypothesis of a uniform areal density of vortices, this total number  of
vortex pierce points on the combined surface $\Sigma + S$ is also given by
\begin{equation}
\label{1plus22}
N_{tot}= \frac{\hat{p}(A_{\Sigma}+A_S)}{\lambda^2}.
\end{equation}
Combining these equations, one finds
\begin{equation}
\label{combine}
N_I=\frac{\hat{p}(A_S-A_{\Sigma})}{\lambda^2}.
\end{equation}

To calculate the area law in the DGA, we can easily use the formulas of
Sec.~\ref{principles} for the flat $\Sigma$ surface, which has a total of
$N_{\Sigma}=A_{\Sigma}/\lambda^2$ $\lambda$-squares:
\begin{equation}
\label{sigma}
\langle W \rangle = [\bar{p}-\hat{p}]^{N_{\Sigma}}\rightarrow e^{-K_F
A_{\Sigma}},
\end{equation} 
with $K_F  =2\hat{p}/\lambda^2$ as before.  It is more interesting to
calculate it from the point of view of the other spanning surface $S$.  Here
we must account for the diminished probability of occupation of this surface
by linked vortices, since some of them, as counted by $N_U$, are inert.  So
we change the probabilities by adding to $\bar{p}$, the probability of no
occupation, the probability $\hat{p}_I\equiv N_I/N_S$ of occupation by an inert
vortex of Type III.  This gives
\begin{equation}
\label{punlink}
\bar{p}\rightarrow
\tilde{p}_0=\bar{p}+\hat{p}_I=\bar{p}+\hat{p}(1-\frac{A_{\Sigma}}{A_S})=
1-\frac{\hat{p}A_{\Sigma}}{A_S}.
\end{equation}
Similarly, the new link probability $\tilde{p}\equiv 1-\tilde{p}_0$ is
\begin{equation}
\label{plink}
\tilde{p}=\frac{\hat{p}A_{\Sigma}}{A_S}.
\end{equation}
Using Eq.~(\ref{sigma}) with the new probabilities for the surface $S$ one
gets,  going to the DGA limit as before:
\begin{equation}
\label{newsurf}
\langle W \rangle = [\tilde{p}_0-\tilde{p}]^{N_S}\rightarrow
e^{-2(\tilde{p}/\lambda^2 )A_S} = e^{-2(\hat{p}/\lambda^2 )A_{\Sigma}}.
\end{equation}

\section{\label{2loops} Compound Wilson loops} 

\begin{figure}
\includegraphics[height=3in]{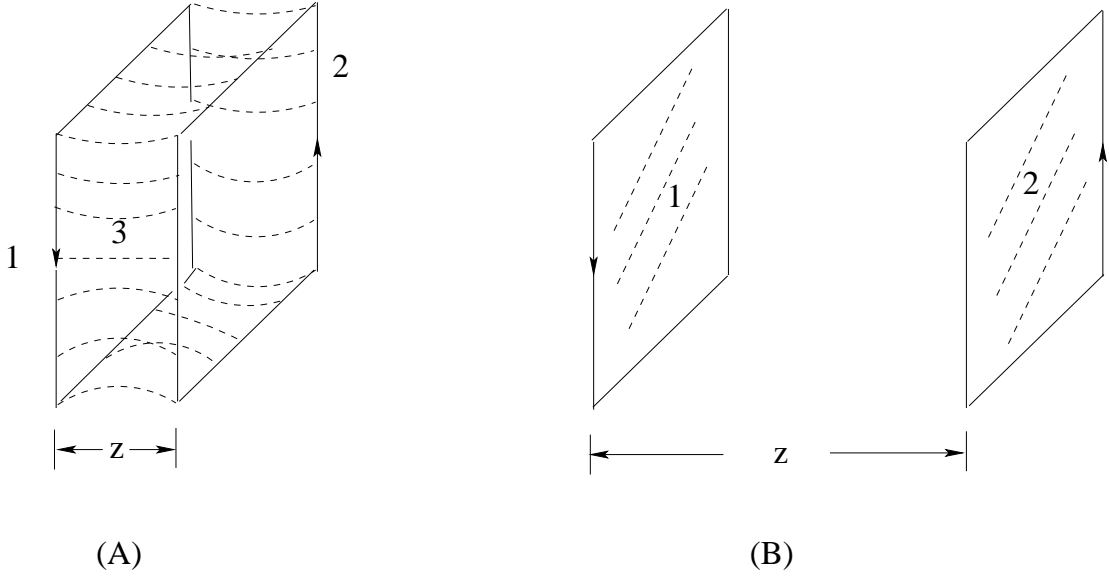}
\caption{\label{fig3} Two identical but oppositely-oriented rectangular Wilson
loops 1 and 2, separated by a distance $z$.  (A) When $z$ is small compared to
loop scales, the Wilson-loop area, labeled 3, connects the loops.  (B)  When
$z$ is large, the Wilson-loop areas are the disjoint areas, labeled 1 and 2, 
of each loop.}
\end{figure}

 Consider the compound Wilson loop shown $W_{comp}$ in Fig.~\ref{fig3},
composed
of two identical but oppositely-oriented Wilson loops.  [The orientation is
irrelevant in $SU(2)$.]  They are  separated transversely by a distance $z$. 
The relevant expectation value is 
\begin{equation}
\label{comploop}
\langle W_{comp} \rangle = \langle W(1) W^{\dagger}(2) \rangle
\end{equation}

When the distance $z$ between the loops is small compared to the loop,
intuition \cite{corn2004} suggests that the  surface whose area should appear
in
the area law, as in Eq.~(\ref{wloop}), is the minimal surface 3 joining the
two loops as shown in  Fig.~\ref{fig3}(A).  When $z$ is large, intuition
suggests that the situation in Fig.~\ref{fig3}(B) holds, where each loop is
spanned by its own minimal surface with no connection to the other surface. 
We will show that intuition is indeed correct for the compound area law of
center vortex theory, and give an approximate interpolation formula for
intermediate values of $z$. 

Begin with the $SU(2)$ case, where the orientation of the Wilson loop does not
matter, and $W=W^{\dagger}$.  As usual, we assume that the time extent $T$,
the spatial extent $R$ of the Wilson loop, and their separation $z$ are all
large compared to the QCD scale length $\lambda$ and assume that $T\gg R$.  In
this limit the spanning surface $S_3$ of Fig.~\ref{fig3}(A) is nearly flat,
and has area $A_3\simeq 2zT$.  (We ignore the contribution $\simeq 2zR$ from
the top and bottom.)  The Wilson loop surfaces $S_{1,2}$ have areas
$A_1=A_2=RT$.  

Imagine now a configuration where all surfaces $S_{1,2,3}$ exist, so that there
is a closed surface with two marked contours, the Wilson loops 1 and 2.  This
constitutes a minor generalization of the configuration already considered in
Sec.~\ref{unlink}.  There are several ways that vortices can be linked or
inert (in the sense of Sec.~\ref{types}), after the renormalization of Type
I and Type II vortices.  Use the notation $N_i$ for the total number of
vortices penetrating surface $S_i$.  These obey
\begin{equation}
\label{2loopn}
N_i=\frac{\hat{p}A_i}{\lambda^2}.
\end{equation}
Each of these is subdivided as follows:  The number of vortices piercing
$S_{1}$ and $S_3$ an odd number of times is called $N_{13}$, with analogous
notation for $N_{23}=N_{13}$; the number piercing surface $S_1$ and $S_2$ each
an odd number of times is $N_{12}$; and the number entering $S_3$ and then
exiting the same surface is $N_{33}$.

As indicated in Sec.~\ref{probsurface}, the probability that a vortex known to
be at a point remote from a surface that then penetrates the surface (once) is
proportional to the surface area, and is finite when the QCD scale length
$\lambda$ vanishes.  If we require the number of vortices penetrating one
surface, say $S_1$, and then another surface, say $S_2$, that probability is
proportional to $A_1A_2$.  So we now assume that
\begin{equation}
\label{nassump}
N_{12}\sim A_1A_2=A_1^2;\;\;N_{13}\sim A_1A_3;\;\;N_{33}\sim A_3^2.
\end{equation} 
These vortex-linking numbers are related to the total vortex-piercing numbers
by
\begin{equation}
\label{sumrule}
N_1=N_{12}+N_{13};\;\;N _3=2N_{13}+N_{33}.
\end{equation}
Eqs.~(\ref{2loopn},\ref{nassump},\ref{sumrule}) are easily solved to yield
\begin{equation}
\label{solve2}
N_{12}=\frac{2\hat{p}A_1^2}{\lambda^2(2A_1+A_3)};\;\;N_{13}=\frac{\hat{p}A_1A_3}
{\lambda^2(2A_1+A_3)};\;\;N_{33}=\frac{\hat{p}A_3^2}{\lambda^2(2A_1+A_3)}.
\end{equation}

We will now compute the expectation value $\langle W \rangle $ of the product
$W\equiv W_1W_2$ of the two Wilson loops, from the point of view of the
surfaces $S_{1,2}$.    
Only the number $N_{13}$ contributes non-trivially to an $SU(2)$ Wilson loop. 
As in Sec.~\ref{unlink} we introduce modified probabilities
\begin{equation}
\label{modprob}
\tilde{p}_1\equiv
\frac{\lambda^2 N_{13}}{A_1}=\frac{\hat{p}A_3}{2A_1+A_3};\;\;\tilde{p}_0
=1-\tilde{p}.
\end{equation}
Just as in calculating the Wilson loop VEV from Eq.~(\ref{newsurf}), we have in
the DGA:
\begin{equation}
\label{newsurf2}
\langle W_{comp} \rangle = (\tilde{p}_0-\tilde{p})^{2A_1/\lambda^2} \rightarrow
\exp -[\frac{2K_FA_1A_3}{2A_1+A_3}]=\exp -[\frac{2K_FTRz}{R+z}].
\end{equation}

This formula is only approximate, but it shows features that we believe are
generally correct (and one feature that is not correct).  For example, the
heavy-quark potential $V$ [coefficient of $-T$ in the exponent of
Eq.~(\ref{newsurf2})] has the behavior $V\simeq 2K_Fz$ in the limit $z\ll R$,
showing that the the area law of two $SU(2)$ Wilson loops [or, as in
Fig.~\ref{fig3}, two oppositely-directed Wilson loops] disappears as the two
loops
approach each other and form an $N$-ality zero configuration.  (For $N$-ality
zero loops there is a pseudo-area law, coming from the finite size of vortices
\cite{corn83}, at distances  $z\sim \lambda$, but that is irrelevant here.) 
In the opposite limit of $z\gg R$ we find $V=2K_FR$, so the VEV is just the
product of the separate VEVs for two Wilson loops.  While this in itself is
correct, the approach to this limit cannot be, for it would yield a residual
potential $V-2K_FR$ which vanishes only like $1/z$.  In actuality, at some
point when $R\simeq z$ the spanning surface switches on the length scale
$\lambda$ from $S_3$ to $S_1+S_2$, much as the corresponding soap-bubble
surface would for two physical wire loops 1 and 2.  We do not see this
breaking because we have not included effects coming from a network of gluon
world lines running from loop 1 to loop 1, from loop 2 to loop 2, and from
loop 1 to loop 2.  The simplest step in the formation of this network has been
discussed \cite{corn05} in connection with baryonic and mesonic hybrids,
having extra gluons along with their quark content.  There it is shown that a
single gluon acts like a physical string which separates a Wilson-loop
spanning surface into two surfaces, and the string requires extra energy to be
stretched if the string stretching leaves the minimal spanning surface.  The
fluctuations associated with this stretching should yield the L\"uscher
\cite{luscher} term, as well as leading to the transition from surface $S_3$
to $S_1+S_2$ as the gluons in the network recombine.  But when this stretching
is not too great the potential should behave roughly as given in
Eq.~(\ref{newsurf2}).

In $SU(N)$ with $N>2$ the orientation of the Wilson loops does matter. If the
loops have opposite orientation, as shown in Fig.~\ref{fig3}, the problem is
essentially the same as for $SU(2)$.  But one should also consider the problem
of two loops of the same orientation.  Since for $SU(3)$ the treatment of this
problem is quite similar to that for the baryon, which is a compound of three
Wilson loops, we defer further discussion to Sec.~\ref{baryon}.

\section{\label{baryon} Vortices analogous to inert vortices and the baryonic
area law}

\begin{figure}
\includegraphics[height=3in]{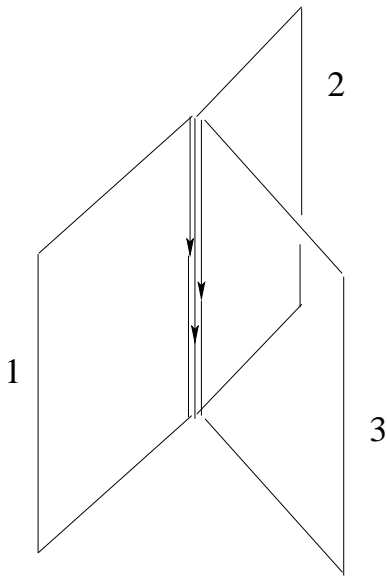}
\caption{\label{fig4} A baryonic Wilson loop in $SU(3)$ is composed of three
simple Wilson loops sharing a common central line (expanded in the figure). 
The central line is invisible to $SU(3)$ center vortices.}
\end{figure}

It is by now well-established both in center vortex theory \cite{cornbar} and
on the lattice \cite{suganuma,forcrand} that, as shown in Fig.~\ref{fig4}, in
$SU(3)$ the heavy-quark baryonic
potential has three Wilson-loop surfaces spanning a boundary consisting of the
quark world lines, with the three lines coinciding along a central line (the
so-called Y law).   
This is a three-fold compound Wilson loop, sharing some of the features of the
double loop discussed in Sec.~\ref{2loops}.  In particular, a vortex may
pierce two loops.   Such a vortex  is not inert, as were the vortices piercing
surfaces 1 and 2 of Fig.~\ref{fig3} and discussed in the previous section. 
Instead, linkage with these analogs of inert vortices are important to
establish that the string tension in the baryonic area law is precisely the
mesonic string tension $K_F$.

The analysis of the dilution factor $\alpha$ for $SU(3)$ is slightly more
complicated, because a
vortex piercing twice is not unlinked as it is in $SU(2)$; it is simply
equivalent to an antivortex piercing once.  Given the complete symmetry
between vortex and antivortex, this means that an antivortex piercing twice is
equivalent to a vortex piercing once.  There is a modification of the
numerical value of the dilution factor, since it is now possible to dilute
unit link number by occupying a minimum of two sites, rather than a minimum of
three as for $SU(2)$.  However, the string tension is still lessened by a
single dilution factor $\alpha$, just as for $SU(2)$, and we can continue to
use the notation developed for that case.  We will not pursue the question of
what the value of the dilution factor is for $SU(3)$; it must be quite similar
to that for $SU(2)$.

The $SU(3)$ version of the standard area law for a single Wilson loop, given in
Eq.~(\ref{vev}) for $SU(2)$, is
\begin{equation}
\label{su3law}
\langle W \rangle =\{\bar{p}+\frac{\hat{p}}{2}[e^{2\pi i/3}+e^{-2\pi
i/3}]\}^{A_S/\lambda^2}=[1-\frac{3\hat{p}}{2}]^{A_S/\lambda^2}.
\end{equation}
Here we use $\hat{p}/2$ as the probability (accounting for dilution) that a
vortex of flux $2\pi /3$ pierces the surface, with equal probability that the
antivortex of flux $-2\pi /3$ pierces it; this means that, as before,
$\bar{p}=1-\hat{p}$.  This leads immediately to the DGA $SU(3)$ string tension
$K_F=3\hat{p}\lambda^{-2} /2$.

\begin{figure}
\includegraphics[height=3in]{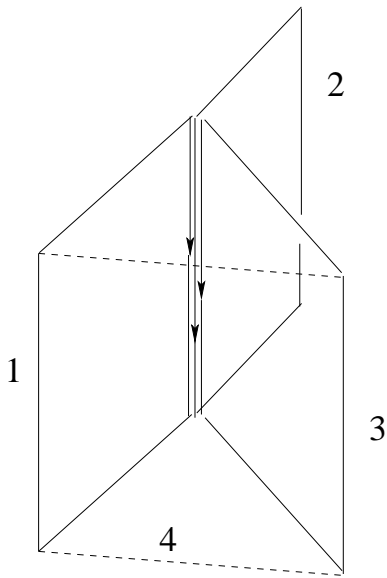}
\caption{\label{fig5} We introduce a new surface labeled 4 that creates a
closed surface consisting of the two Wilson-loop surfaces for quarks 1 and 3
plus a surface joining these.}
\end{figure}

To discuss the baryonic area law in similar terms to those used with the double
Wilson loop, introduce a surface made of flat pieces, labeled 4 and shown in
Fig.~\ref{fig5}, that forms a closed surface when combined with the surfaces
spanning Wilson loops 1 and 3.

In what follows we will be careful to distinguish vortices from antivortices,
although at the end the symmetry between them makes this simply a distinction
for convenience of exposition.
As in Sec.~\ref{2loops} we introduce the number $N_{13}$ as the number of times
that the same vortex (not antivortex) pierces both spanning surface 1 and
surface 3 an odd number of times, and $N_{14}$ is the number of times a vortex
pierces both surface 1 and 4.  We consider that the dilution factor $\alpha$
has been applied, so we can take this odd number to be just one.  The number
$N_1$ is the number of  vortex piercings on surface 1, and  it obeys
\begin{equation}
\label{baryonpierce}
N_1=\frac{\hat{p}A_1}{2\lambda^2}=N_{13}+N_{14}.
\end{equation} 
The factor of 2 in the denominator arises because there are just as many
antivortices piercing any given surface; it is the same factor of two dividing
the explicit probabilities in the area law of Eq.~(\ref{su3law}).

Any vortex that pierces surface 1 once must leave through either surface 3 or
surface 4.  If it leaves through surface 4 it is associated with a phase
factor $\exp [2\pi i/3]$; if it leaves through surface 3 it is associated with
a phase factor $\exp [-2\pi i/3]$.  For any such vortex configuration there is
an equally probable antivortex configuration, with the phase factors
interchanged.  So the {\em total} number of piercing, vortices plus
antivortices, is $2N_1$, and the number of antivortices piercing surface 3 is
$N_{13}$, etc.  This, plus the relation of Eq.~(\ref{baryonpierce}), means
that the standard calculation of  the baryonic area law using surface 1 (plus
surfaces 2 and 3), giving the Y law, is exactly the same as accounting for the
linkages of vortices and antivortices through surfaces 1 and 4, plus linkages
through surfaces 1 and 3 (plus permutations for the other two quarks).  There
really are no unlinked Type III vortices in this problem, the way there were
in Secs.~\ref{unlink}, \ref{2loops}.  The baryonic string tension is precisely
the mesonic string tension, because when it is calculated from the three quark
surfaces 1, 2, and 3 separately, without regard to the other two quarks, it is
just the standard simple Wilson loop calculation.  But it is just the same
when linkages to two quarks are considered. Of course, no one doubted the
equality of string tensions for baryons and mesons, but the point was to see
how it worked out in the center vortex picture.

\section{\label{conc}  Summary and conclusions}

We have discussed semi-quantitatively some of the effects of inert vortices
that do not couple as effectively as they might to a Wilson loop, and that
change the probability of linkage by a dilution factor $\alpha\leq 1$.  In
turn, this
changes the $SU(2)$  DGA string tension from $K_F=2\rho$ to $K_F=2\alpha \rho$,
where $\rho$ is the areal density of vortices piercing a large flat surface.
  Our estimate, based on partial sums of non-backtracking walks, is $\alpha =
0.6\pm 0.1$.
Understanding inert vortices also
leads to understanding of how it is that there is a unique area in the area
law, even though any surface spanning a Wilson loop is suitable for counting
link numbers in the center vortex picture.  We found a finite-range Van der
Waals force, due to inert vortices, that links two mesonic Wilson loops, in
this case vortices that are simultaneously linked to both Wilson loops.  And
finally we showed that the analogs of inert vortices in the two-loop problem do
not interfere with the usual
formulation of the area law for $SU(3)$ baryons, based on considering the three
Wilson loops as independent.  

There seems to be no possibility of a detailed analytic approach to these
problems, which therefore are best studied further with lattice computations. 
One can, of course, create a center vortex condensate through simulation of
the underlying non-Abelian gauge theory, but it would also be very interesting
and possibly simpler to study inert-vortex effects with simulations that begin
with an {\em a priori} vortex model, similar in spirit to the approach of
\cite{engrein}, rather than having to deal with all the
complications of the full gauge theory.  Indeed, it seems that relating $K_F$
and $\rho$ on the lattice has numerous complications \cite{bfgo,forp,ltmr},
and has not been attempted recently.  But the simulation of two Wilson loops
seems approachable on the lattice.

\newpage

\appendix

\section{ \label{probsurface} Probabilities  issues}

This appendix discusses two issues:  1)  A brief review of well-known analysis
for return probabilities for random walks with backtracking allowed; 2) The
probability for a vortex to pierce two separated Wilson loops.  The first
serves for cautionary notes in the problem calculating the dilution factor of
Sec.~\ref{wloops}, and
the second arises in finding the area law for the compound Wilson loop of 
Sec.~\ref{2loops}.  For the most part and for dimension $d\leq 4$, analytic
results are only available for random walks without self-avoidance
constraints, and we will only discuss those in the next subsection.

\subsection{The gambler's ruin problem
[Sec.~\ref{wloops}]}

The only problems that one has any real chance to analyze are for
non-self-avoiding walks, and we will review a famous one, the gambler's ruin
problem, here.  The basic probability concepts are the same for self-avoiding
walks, and so this review should be helpful to those unfamiliar with the
underlying ideas.  The cautionary note here is that while the $d=1$ gambler's
ruin problem seems superficially quite similar to those of Sec.~\ref{wloops},
in practice they are very different, both because the questions of
Sec.~\ref{wloops} have three-dimensional effects and because they deal with
self-avoiding walks.

We ask for the probability $q_1(K;m)$ that an unbiased random walk starting
from the origin
on a lattice of points in $d=1$ 
will return to  the point $m\geq 0$ for the first
time after $K$ steps. The case $m=0$ is the standard gambler's ruin problem. 
For an unbiased walk (all step probabilities equal
to $1/(2d)$ in $d$ dimensions) P\'olya long ago proved that in the limit of
infinite steps the probability of return to the origin is unity in $d=1,2$ but
less than one in all
other dimensions.  This is perhaps plausible from the fact that the probability
of being at the origin after $N$ steps in $d$ dimensions is, for $N\gg 1$,
\begin{equation}
\label{app1}
p_d(N) \sim  N^{-d/2},
\end{equation}
and the sum over $N$ diverges at large $N$ in $d=1,2$ but not in higher
dimensions.

 As in Sec.~\ref{wloops}, the
relation between the $p_1(N;m)$ and the $q_1(N;m)$ is
\begin{equation}
\label{feller}
p_1(N;m)=\sum_{J=1}^Np_1(N-J;m=0)q_1(J;m)\;\;\;\; \;(p_1(N=0;m)=\delta_{m,0}).
\end{equation}
Note that in the sum on the right $p_1$ has $m$ set equal to zero, because we
are compounding the probability of first return to $m$, as given by $q_1$,
with the probability of being at this same point $m$ after more steps.

To solve these relations, define the generating functions   $P_1(s;m)$ and
$Q_1(s;m)$:
\begin{equation}
\label{genfunct}
P_1(s;m)=\sum_0 p_1(N;m)s^N;\;\;Q_1(s;m)=\sum_2q_1(N;m)s^N.
\end{equation}
Eq.~(\ref{feller}) then translates to
\begin{equation}
\label{fellereq}
P_1(s;m)=\delta_{m,0}+Q_1(s;m)P_1(s;0).
\end{equation}

It is straightforward to find $p_1(N;m)$ and $P_1(s;m)$.
The probability $p_1(N;m)$ is the standard random-walk probability, given by
\cite{montroll}
\begin{equation}
\label{montroll1}
p_1(N;m)=\frac{1}{2\pi }    \int_0^{2\pi}d\theta
[\cos \theta]^N\exp [im\theta] .
\end{equation}  
From this we find the generating function (for $m\geq 0$)
\begin{equation}
\label{meqn}
 P_1(s;m)=\frac{1}{2\pi}\int_0^{2\pi}d\theta \frac{e^{im\theta}}{1-s\cos
\theta}=\{\frac{1}{s}[1-(1-s^2)^{1/2}]\}^m(1-s^2)^{-1/2},
\end{equation}
and the gambler's ruin ($m=0$) generating function is 
 \begin{equation}
\label{generate}
P_1(s;m=0)=[1-s^2]^{-1/2}.
\end{equation}
 This yields for the gambler's ruin problem the well-known result
\begin{equation}
\label{d1q}
Q_1(s;m=0)=1-P_1(s;m=0)^{-1}=1-[1-s^2]^{1/2}.
\end{equation}
 One learns from this that the probabilities for first return after $K$= 2, 4,
6,
8,$\dots$ steps
are 1/2, 1/8, 1/16, 5/128,$\dots$ independent of the total number of steps in
the random walk (if this number is larger than $K$), and that
the probability of ever returning is unity.  This follows from
Eqs.~(\ref{app1}, \ref{genfunct}), which shows that $P_1(1)$ diverges in $d=1$
(and
also in $d=2$) but
not in higher dimensions. 

For non-zero $m$ we have 
\begin{equation}
\label{mgenfunct}
P_1(s;m)=Q_1(s;M)P_1(s;m=0).
\end{equation}
  It then follows from Eqs.~(\ref{meqn}, \ref{generate}, \ref{mgenfunct})that
\begin{equation}
\label{qgenfunct}
Q_1(s;m)=\{\frac{1}{s}[1-(1-s^2)^{1/2}]\}^m.
\end{equation}
By expanding in $s$, one sees that the  probabilities vanish for
$N<m$, as expected, and by setting $s=1$ in $Q_1(s;m)$ one sees that the 
probability of ever reaching $m$ is unity.  The probability $q_1(N;m)$
peaks for $N\sim m^2$, when $m$ is comparable to the vortex radius of
gyration.

Now consider calculating  the probability of first return anywhere on an
infinite plane in $d=3$.  
The probability
$p_3(N;\vec{m})$ of going from
the origin to lattice point $\vec{m}$ in $N$ steps on an infinite lattice is
given in Eq.~(\ref{3dprob}   and the corresponding generating function is
\begin{equation}
\label{3dgen}
P_3(s;\vec{m})=\frac{1}{(2\pi )^3 }    [\prod_{j=1}^3\int_0^{2\pi}d\theta_j]
\{1-\frac{s}{3}[\cos \theta_1+\cos \theta_2+\cos \theta_3]\}^{-1}\exp
[i\vec{\theta}\cdot\vec{m}];\;P_3(0;\vec{m})=\delta_{\vec{m},\vec{0}}.
\end{equation}
 Take $m_3=0$ in Eq.~(\ref{3dgen}) and sum from
$-\infty$ to $\infty$ over $m_{1,2}$.  The result is, not unexpectedly, a
simple variant on the $d=1$ gambler's ruin problem, and yields
\begin{equation}
\label{summ}
P_3(s)\equiv \sum_{m_1,m_2=-\infty}^{\infty}P_3(s;m_1,m_2,m_3=0)=
\frac{1}{\pi}\int_0^{\pi}\frac{d\theta}{1-\frac{2s}{3}-\frac{s}{3}\cos \theta}
=[1-\frac{4s}{3}+\frac{s^2}{3}]^{-1/2}.
\end{equation}
Using a standard expansion and Eq.~(\ref{d1q}) we find the probabilities
$q_3(K)$:
\begin{equation}
\label{expand}
q_3(K)=-3^{-K/2}C_K^{-1/2}(\frac{2}{\sqrt{3}})
\end{equation}
where the $C_K^{-1/2}$ are Gegenbauer polynomials.

\subsection{\label{compoundloops} Probability that a vortex pierces two
separated
Wilson loops [Sec.~\ref{2loops}]}

The next problem comes up in the compound Wilson loop estimates of
Sec.~\ref{2loops}, where a vortex can pierce two separated Wilson loops.
In $d=3$, consider the  plane surface of dimensions $L_1,L_2$ centered on and
perpendicular to the $z$-axis at a distance $M$ along this axis from the
origin.  We ask for the probabilities $p_3(N;L_1,L_2,M)$ to end up on this
surface after $N$ steps; the probability $q_3(N;L_1,L_2,M)$ of reaching this
surface for the first time after $N$ steps; and the probability of ever
reaching it.  Also to be calculated are the corresponding generating functions
$P_3(s;L_i,M)$ and $Q_3(s;L_i,M)$.  We assume that all lengths $L_i,M$ are
large in lattice units, so that the number of steps is also large.

The first probability is a sum over the surface of the probability given in
Eq.~(\ref{3dprob}):
\begin{equation}
\label{p3llm}
p_3(N;L_i,M)=\sum_{m_i=-L_i/2}^{m_i=L_i/2}p_3(N;\vec{m})|_{m_3=M}. 
\end{equation}
The sum is easily done to yield
\begin{equation}
\label{p3llm2}
p_3(N;L_i,M)=\frac{1}{(2\pi )^3 3^N}    [\prod_{j=1}^3\int_0^{2\pi}d\theta_j]
[\cos \theta_1+\cos \theta_2+\cos \theta_3]^N\exp [iM\theta_3]\frac{\sin
[(L_1+1)\theta_1 /2]}{\sin [\theta_1 /2]}\frac{\sin [(L_1+1)\theta_2 /2]}{\sin
[\theta_2 /2]}.
\end{equation}
In the large-$N$ limit $N\gg L_i^2$, one finds an area factor emerging:
\begin{equation}
\label{p3llma}
p_3(N;L_i,M)\simeq \frac{A}{(2\pi )^3 3^N}   
[\prod_{j=1}^3\int_0^{2\pi}d\theta_j]
[\cos \theta_1+\cos \theta_2+\cos \theta_3]^N\exp [iM\theta_3]
\end{equation}
where $A=L_1L_2$ is the area of the surface.  This happens because the
integrand is only appreciable when $\theta_i\leq N^{-1/2}$.
The generating function for this probability, if needed, is constructed as in
Eq.~(\ref{3dgen}), including summing over $m_1,m_2$.

 The probability $q_3(N;L_1,L_2,M)$ is determined by an analog of
Eq.~(\ref{feller}), including a sum over surface variables:
\begin{equation}
\label{feller1}
p_3(N;L_i,M)=\sum_J\sum_{m_i}q_3(J;m_i,M)p_3(N-J;m_i,0)
\end{equation}
where the sum over $m_1,m_2$ is delimited as in Eq.~(\ref{p3llm}).
To see what this sum means, write the $q$-probability in Fourier form
\begin{equation}
\label{qfourier}
q_3(J;m_i,M)=\frac{1}{(2\pi )^3}\int d^3\theta \tilde{q}_3(J;\vec{\theta})
e^{i\vec{\theta}\cdot \vec{m}}
\end{equation}
with $m_3=M$.
We find, using Eq.~(\ref{3dprob}),
\begin{equation}
\label{qfourier1}
p_3(N;L_i,M)= \sum_J\int \frac{d^3\theta}{(2\pi )^3}e^{i\theta_3M}
\tilde{q}_3(J;\vec{\theta})
\prod_{j=1,2}\frac{\sin
[L_j(\theta_j+\alpha_j+1)/2]}{\sin[(\theta_j+\alpha_j+1/2]} 
\int \frac{d^3\alpha}{(2\pi )^3}[(\sum \cos \alpha_i)/3]^{N-J}.
\end{equation}

We anticipate, and can confirm later, that in the limit $N\rightarrow \infty$
the $\theta_j$ are of order $J^{-1/2}$ and the $\alpha_j$ are of order
$(N-J)^{-1/2}$.  It turns out that $N-J$ is large compared to $J$ (which is of
order $M^2$) and so we can drop the $\alpha_{1,2}$ in the argument of the sine
functions in Eq.~(\ref{qfourier1}).  The $\alpha$ integral then factors out,
and is given by the probability $p_3(N-J;\vec{0})$ of returning to a given
point after $N-J$ steps.  

The problem is now solved, in principle, by using generating functions, as in
Eq.~(\ref{mgenfunct}).  Or one can study the sum of Eq.~(\ref{qfourier1})
directly, and find by a scaling argument that the maximum value of the
$q$-probability on the right-hand side behaves like $A/J\sim A/M^2$.  Note that
when lattice lengths $L_i,M$ are converted to physical lengths by multiplying
by $\lambda$ this probability remains finite.


\newpage

\begin{thebibliography}{99}
\bibitem{engrein} M.~Engelhardt and H.~Reinhardt, Nucl.\ Phys.\ B {\bf 585},
591 (2000); M.~Engelhardt, M.~Quandt, and H.~Reinhardt, Nucl. Phys. {\bf
B685}, 227 (2004); M.~Quandt, H.~Reinhardt, and M.~Engelhardt, Phys. Rev. D
{\bf 71}, 054026 (2005).
\bibitem{greensite} J.~Greensite, Prog. Part. Nucl. Phys. {\bf 51}, 1 (2003).
\bibitem{bfgo} L. Del Debbio, M. Faber, J. Giedt, J. Greensite, and \u{S}
Olejn\'ik, Phys. Rev. D {\bf 58}, 094501 (1998); R. Bertle, M. Faber, J.
Greensite, and  \u{S} Olejn\'ik, JHEP {\bf 0010}, 007 (2000); M. Faber, J.
Greensite, and \u{S} Olejn\'ik, JHEP {\bf 0111}, 053 (2001).
\bibitem{forp} P. de Forcrand and M. Pepe, Nucl. Phys. B {\bf 598}, 557 (2001).
\bibitem{ltmr} K. Langfeld, O. Tennert, M. Engelhardt, and H. Reinhardt, Phys.
Lett. B {\bf 452}, 301 (1999).
\bibitem{corn2004} J.~M.~Cornwall, Phys.\ Rev.\ D {\bf 69}, 065019 (2004).
\bibitem{cornbar} J.~M.~ Cornwall, Phys. Rev. D {\bf 69}, 065013 (2004).
%\bibitem{corn041}  J.~M.~Cornwall, Phys.\ Rev.\ D {\bf 69}, 065019 (2004).
\bibitem{nemir} A.~M.~Nemirovsky and M.~D.~Coutinho-Filho, Phys.\ Rev.\ A{\bf
39}, 3120 (1989).
\bibitem{tempfish}  H.~N.~V.~Temperley and M.~E.~Fisher, Phil.\ Mag.\ {\bf 6},
1061 (1961); M.~E.~Fisher, Phys.\ Rev.\ {\bf 124}, 1664 (1961); P.~W.~Kastelyn,
J.\ Math.\ Phys. {\bf 4}, 287 (1961).
\bibitem{montroll} E.~W.~Montroll, J.~SIAM {\bf 4}, 241 (1956).
\bibitem{feller}  W. Feller, {\em An introduction to probability theory and its
applications}, Wiley, New York, 1951.
\bibitem{corn83} J.~M.~Cornwall, {\it Workshop on 
Non-Perturbative Quantum Chromodynamics}, ed.~K.~A.~Milton and A.~Samuel
(Birkh\"auser, Boston, 1982), p.~119; Phys.\ Rev.\
D{\bf 57}, 7589 (1998).
\bibitem{corn05} J.~M.~Cornwall, Phys.\ Rev.\ D {\bf 71}, 056002 (2005).
\bibitem{luscher} M.~L\"uscher, Nucl.~Phys.~{\bf B180}, 317 (1981);
M.~L\"uscher, K.~Symanzik, and P.~Weisz, Nucl.\ Phys.\ {\bf B173}, 365 (1980).
\bibitem{suganuma} T. T. Takahashi, H. Matsufuru, Y. Nemoto, and H. Suganuma, 
Phys. Rev. Lett. {\bf 86}, 18 (2001); Phys. Rev. D{\bf 65}, 114509 (2002).
\bibitem{forcrand}  C. Alexandrou, P. de Forcrand, and O. Jahn, Mucl. Phys. B
(Proc. Suppl.) {\bf 119}, 667 (2003). 
\end{thebibliography}
\end{document}